\documentclass[12pt]{article}
\usepackage{jheppub}
\renewcommand{\theequation}{\thesection.\arabic{equation}}

\newcommand{\ba}{\begin{array}}
\newcommand{\ea}{\end{array}}
\newcommand{\beq}{\begin{equation}}
\newcommand{\eeq}{\end{equation}}
\newcommand{\bea}{\begin{eqnarray}}
\newcommand{\eea}{\end{eqnarray}}

\def\m{\mu}

\def\bce{\begin{center}}
\def\ece{\end{center}}

\def\nonu{\nonumber}

\def\pa{\partial}
\def\al{\alpha}
\def\be{\beta}

\def\ep{\epsilon}

\def\th{\theta}

\def\la{\lambda}
\def\La{\Lambda}

\def\eps6{{\displaystyle \mathop{\epsilon}^{6}}{}}
\def\g6{{\displaystyle \mathop{g}^{6}}{}}

\def\nab6{{\displaystyle \mathop{\nabla}^{6}}{}}
\def\to{\rightarrow}

\def\0{{\sst{(0)}}}
\def\1{{\sst{(1)}}}
\def\2{{\sst{(2)}}}
\def\3{{\sst{(3)}}}
\def\4{{\sst{(4)}}}
\def\5{{\sst{(5)}}}
\def\6{{\sst{(6)}}}
\def\7{{\sst{(7)}}}
\def\8{{\sst{(8)}}}

\def\ba{\begin{array}}
\def\ea{\end{array}}
\def\beq{\begin{equation}}
\def\eeq{\end{equation}}
\def\be{\begin{equation}}
\def\ee{\end{equation}}

\def\to{\rightarrow}

\def\la{\lambda}
\def\eps{\epsilon}

\def\a{{\alpha}}
\def\b{{\beta}}

\def\th{{\theta}}

\def\ba{\begin{array}}
\def\ea{\end{array}}
\def\beq{\begin{equation}}
\def\eeq{\end{equation}}
\def\be{\begin{equation}}
\def\ee{\end{equation}}

\def\to{\rightarrow}

\def\la{\lambda}
\def\eps{\epsilon}

\def\a{{\alpha}}
\def\b{{\beta}}

\def\th{{\theta}}

\def\eps6{{\displaystyle \mathop{\epsilon}^{6}}{}}

\def\nab6{{\displaystyle \mathop{\nabla}^{6}}{}}

\newcommand{\bean}{\begin{eqnarray*}}
\newcommand{\eean}{\end{eqnarray*}}

\begin{document}
\thispagestyle{empty} \addtocounter{page}{-1}
   \begin{flushright}
\end{flushright}

\vspace*{1.3cm}
  
\centerline{ \Large \bf   
Chiral Algebras of Two-Dimensional 
SYK Models} 
\vspace*{1.5cm}
\centerline{\bf Changhyun Ahn$^{a,b}$
\footnote{On leave from the Department of Physics, Kyungpook National University, Taegu
  41566, Korea and 
address until Aug. 31, 2018:
C.N. Yang Institute for Theoretical Physics,
Stony Brook University,
Stony Brook, NY 11794-3840, USA
}
    and Cheng Peng$^c$} 
\vspace*{1.0cm} 
\centerline{\it
$^a$  C.N. Yang Institute for Theoretical Physics,
Stony Brook University,
Stony Brook, NY 11794-3840, USA}
 \centerline{\it 
$^b$ Department of Physics, Kyungpook National University, Taegu
41566, Korea} 
\centerline{\it
$^c$ Department of Physics, Brown University, 
Providence, RI 02912, USA}
 \vspace*{1.0cm} 
 \centerline{\tt ahn@knu.ac.kr,
   \qquad cheng$\underline{~}$peng@brown.edu}

\vskip2cm

\centerline{\bf Abstract}
\vspace*{0.5cm}

We study chiral algebras in the $\bar{Q}$-cohomology of two dimensional SYK models with extended supersymmetry.  
In a special limit  
discovered in~\cite{Peng:2018zap}, we are able to construct explicitly a ``vertical" single-particle higher-spin algebra that is bilinear in the fundamental fields. This algebra can be regarded as the counterpart, when going away from criticality, of the  infrared emergent higher-spin symmetry of the $\mathcal{N}=(0,2)$ SYK model. Moreover, a second ``horizontal" single-particle higher-spin algebra 
appears in this limit. Together with the vertical algebra they generate a stringy algebra with a  ``higher spin square" structure that is believed to appear in the tensionless limit of string theory. 
On the other hand, we do not find single-particle higher-spin algebra away from the special limit, which is consistent with the result in~\cite{Peng:2018zap}. 
Our analysis is carried out for each individual realization of the random couplings and for finite $N$ (and $M$), which in particular indicates that the conclusion in \cite{Peng:2018zap} is robust to $1/N$ corrections.

\baselineskip=18pt
\newpage
\renewcommand{\theequation}
{\arabic{section}\mbox{.}\arabic{equation}}

\tableofcontents

\section{Introduction}
The Sachdev-Ye-Kitaev (SYK) model~\cite{Sachdev:1992fk,Parcollet:1997ysb,PG,KitaevTalk1,KitaevTalk2,Maldacena:2016hyu, Kitaev:2017awl} provides new insights into quantum gravity thanks to its  perturbative  solvability~
\cite{KitaevTalk1,Polchinski:2016xgd,Maldacena:2016hyu,Jevicki:2016bwu,Gross:2017hcz, Gross:2017aos,	Kitaev:2017awl,Klebanov:2018fzb,Rosenhaus:2018dtp}. 
The attractive features of the SYK model include its chaotic nature~\cite{KitaevTalk1,KitaevTalk2,Maldacena:2016hyu, Bagrets:2017pwq,Gu:2018jsv}, and its explicit and spontaneous broken reparameterization symmetry in the infrared that results in a simple Schwarzian derivative action~\cite{KitaevTalk2,Bagrets:2016cdf,Stanford:2017thb,Mertens:2017mtv,Jevicki:2016ito, Kitaev:2017awl}. 
These properties are also shared by dilaton gravity systems on nearly AdS$_2$ spacetimes~\cite{Maldacena:2016hyu, Kitaev:2017awl,Strominger:1998yg,Maldacena:1998uz,Almheiri:2014cka, Maldacena:2016upp,Engelsoy:2016xyb, Cvetic:2016eiv,Grumiller:2017qao,Maldacena:2018lmt,Shenker:2013pqa,Shenker:2014cwa,Maldacena:2015waa,Jensen:2016pah,Saad:2018bqo,Kitaev:2018wpr,Yang:2018gdb,Cardenas:2018krd}. Supersymmetric SYK-like models~\cite{Fu2017,Klebanov2017,Peng2017b,Chang:2018sve} are also proposed, see also~\cite{Bulycheva:2018qcp,Peng2017a,Yoon2017} for related discussions.

Besides the gravity sector, there is in addition 
a tower of operators ~\cite{KitaevTalk2,Polchinski:2016xgd,Maldacena:2016hyu,Kitaev:2017awl} with finite anomalous dimensions. As suggested in~\cite{Maldacena:2016hyu}, this tower of operators is analogous to the tower of operators in vector models and a further discussion on this relation can be found in~\cite{Peng:2017kro}. 
Different bulk duals of the tower of  operators are proposed in~\cite{Gross:2017hcz,Gross:2017vhb, Taylor:2017dly,Das:2017hrt,Das:2017wae}.

Higher dimensional generalizations of the SYK model are interesting in making connections to other previously studied models and even to experimental realizations.  There have been  proposals with a discrete spatial direction, see e.g. \cite{Gu:2016oyy}, or with an unconventional kinetic term \cite{Turiaci:2017zwd}, or with marginal irrelevant interactions \cite{Berkooz:2017efq}.

In this paper we study a model, introduced in~\cite{Peng:2018zap}, that is defined in (continuous) 1+1 dimension with canonical kinetic terms and relevant SYK-like random coupling, which is a direct generalization of \cite{Murugan:2017eto}, and see also~\cite{Bulycheva:2018qcp}. The model has an $\mathcal{N}=(0,2)$ supersymmetry. 
In the infrared, this model is dominated by the set of melonic diagrams in the large-$N$ limit and can be solved perturbatively. 
The $\mathcal{N}=(0,2)$ supersymmetry is crucial for our discussion.
The $\mathcal{N}=2$ supersymmetry in the right-moving sector makes the IR solution reliable; the  absence of supersymmetry in the left-moving sector allows a one parameter family of such models. 
In certain limits of this family we observe emergent higher-spin symmetries in the infrared~\cite{Peng:2018zap}.   
This provides an explicit illustration of a connection between  SYK-like models and models with higher-spin symmetry: higher-spin theories can be thought as a subsector of some tensionless limit of string theory~\cite{Sundborg:2000wp,Witten2001Talk,Mikhailov:2002bp,Chang:2012kt,Gaberdiel:2014cha,Vasiliev:2018zer}, while the SYK model should be holographically dual to some string theory with finite tension \cite{Maldacena:2016hyu,Gross:2016kjj}  \footnote{ The model in this paper can be thought of as an appropriate 1+1-dimensional generalization of a similar relation  discussed in \cite{Peng:2017kro}.}.
In addition, it is explicitly observed that the chaotic behaviors disappear as the higher-spin symmetries emerge, which is consistent with the general picture. Furthermore, the pattern of how massive fields becomes conserved higher-spin currents agrees with the pattern of the breaking of the higher spin symmetries from a previously conformal perturbation computation~\cite{Gaberdiel:2015uca}, which provides another evidence to support our claim.  

The field content of this model, namely  $N$ chiral (bosonic) multiplets $\Phi^a$ ($a=1,2, \cdots, N$)
and $M$ Fermi multiplets
$\La^i$ ($i=1,2, \cdots, M$), is identical to those in some two dimensional ${\cal N}=(0,2)$ Landau-Ginzburg models
on flat spacetime~\cite{Witten:1993yc}. The only difference is the form of the interaction: the coupling of the SYK model is \cite{Peng:2017kro} random and the model does not have any global symmetry except for a $U(1)$ charge symmetry. 

In this paper we will discuss a property that is relevant to the $\mathcal{N}=(0,2)$ SYK model, and probably also to the $\mathcal{N}=(0,2)$ Landau-Ginzburg model, which is the chiral algebra of the $\bar{Q}$ cohomology. At  the classical level the quasihomogeneity of the superpotential guarantees~\cite{Witten:1993jg,Silverstein:1994ih} the existence of the conformal stress tensor and implies that the classical action
is invariant under the right-moving $U(1)$ $R$ symmetry. It is easy to verify that the potential of the SYK model is indeed quasi-homogeneous.  At the quantum level, we show that the SYK model satisfies a necessary condition for the existence of the stress tensor. This means there is always a Virasoro subalgebra in the chiral algebra of the 2d SYK model.

In this paper we further study properties of the higher-spin operators, which are those generators of the chiral algebra with spin greater than 2 (and their supersymmetric partners). We would like to understand what is the minimum subalgebra of the chiral algebra that includes at least one, in our case the one with the smallest spin \footnote{There could be other exotic cases where the higher-spin extension has a gap in spin, see e.g. section 5.2.2 of~\cite{Bouwknegt:1992wg} and the reference therein. We do not consider these cases in this work.}, higher-spin operator.  
As we will show in the paper, in the special limit $q\to \frac{N}{M}$, where $q+1$ labels the rank of the interaction of the SYK model~(\ref{J1}), one can identify two different higher-spin subalgebras that are isomorphic to the $\mathcal{W}_{\infty}[\lambda=1]$
algebra studied in \cite{Pope:1990kc}. 
The generators of one of the algebras, which we call ``vertical", can be expressed in terms of single-sum operators that are quadratics of the fundamental fields. Here we denote by ``single-sum" the operators involving one sum of the flavor indices of the chiral or the Fermi multiplet, which is an analogue of the single trace operator in matrix models. The generators of the other algebra can be expressed in terms of single-sum terms that are higher powers of the fundamental fields. The commutators of the two algebra generate a larger algebra that has the structure of a ``higher spin square"~\cite{Gaberdiel:2014cha, Gaberdiel:2015mra}.

One would also ask what is the relation  between  the chiral higher-spin subalgebra of the $\bar{Q}$-cohomology and the chiral higher-spin algebra emerging in the special limit of the SYK model (\ref{J1}) in the infrared \cite{Peng:2018zap}. 
From our analysis in this paper, it is tempting to consider the vertical higher-spin algebra  in the limit $q\to \frac{N}{M}$ ---whose generators are all quadratic in the fundamental fields --- as the counterpart of the emergent higher-spin symmetry discussed in~\cite{Peng:2018zap} when going away from the infrared SYK critical point.

However, once we are away from the limit $q\to \frac{N}{M}$ the inclusion of a single higher-spin operator leads to an infinite dimensional subalgebra that is  larger than a conventional higher-spin algebra: the number of generators of this algebra at each single spin grows as the spin increases, which is in contrast with the conventional higher-spin algebra where there is one operator (supermultiplet) at each spin. This reflects the  stringy nature of this algebra; there is not a simple way to separate out the leading Regge trajectory as opposed to the case in the 0+1 dimensional SYK model. 
This could also be related with the fact that these 2 dimensional SYK models are not maximally chaotic.
Furthermore, the fact that there is no ``single-particle" higher-spin subalgebra in the chiral algebra  away from the $q\to \frac{N}{M}$ limit can be understood holographically as there is not a tower of higher-spin fields in the bulk. Therefore this is consistent with the fact that the Lyapunov exponent of the early time out-of-time-order correlation function is not zero away from this special limit.

\section{Cohomological chiral algebra of 2D supersymmetric SYK model}

\subsection{2D supersymmetric SYK model}
We consider SYK models in continuous 1+1 dimensional spacetime that is of the class discussed in \cite{Murugan:2017eto}.  
Our primary example of this class of model is the one   with $\mathcal{N}=(0,2)$ supersymmetry discussed in \cite{Peng:2018zap}. The model describes $N$ chiral multiplets $\Phi^a$ and $M$ Fermi multiplets $\Lambda^i$ with a random coupling
\be\label{J1}
S_{\rm {int}}=\int d^2x d\theta^+ \Lambda^i J_i(\Phi^a)=\int d^2x d\theta^+ \frac{J_{ia_1\ldots a_q}}{q!} \Lambda^i \Phi^{a_1}\ldots\Phi^{a_q}\ .
\ee
The model with $M=N$ has an enhanced $\mathcal{N}=(2,2)$ supersymmetry  and the model reduces to the one discussed in \cite{Murugan:2017eto}, see also \cite{Bulycheva:2018qcp}. Each superfield contains one bosonic and one fermionic field
\bea
\Phi^a&=&\phi^a + \sqrt{2} \theta^+ \psi^a + 2\theta^+\bar{\theta}^+ \pa_z\phi^a\,,\\
\Lambda^i&=&\lambda^i-\sqrt{2}\theta^+G^i+2\theta^+\bar{\theta}^+\pa_z\lambda^i\,,
\eea
where $z$ is the holomorphic coordinate.

The coupling $J_{ia_1\ldots a_q}$ is randomly chosen from a Gaussian distribution. It is  relevant, with mass dimension one,  so it dominates the physics in the infrared. The IR solution of this model in the limit
\be
M\gg 1\,,\qquad N\gg 1\,,\qquad \mu\equiv\frac{M}{N} \quad {\rm  fixed} \,,
\ee 
is presented in \cite{Peng:2018zap}. 

As shown in \cite{Peng:2018zap}, at generic value of $\m$, this model shares some common features of SYK-like models, such as being  chaotic and the emergence of a conformal symmetry in the IR.
An intriguing feature of this model is  the emergence of higher spin operators in two different limits
\be\label{twolim}
\mu \to \left(\frac{1}{q}\right)^+  \,,\qquad { \rm and } \qquad \mu \to \infty\,,
\ee
 of the model. In each of the two limits one observes a tower of operators that become holomorphic whose left-moving conformal dimension vanish. There is also a tower of antiholomorphic operators in each of the limits as well.
These operators close under the Operator Product Expansion (OPE) and generate a higher-spin symmetry algebra. 
  The appearance of the higher-spin operators and higher-spin symmetry is confirmed by the vanishing of the Lyapunov exponent in the two limits.
  We emphasize that such higher-spin symmetries appear only in the two special limits of the IR model,  which mimic the limit where the string tension approaches to zero.
  
\subsection{Chiral algebra in the $\bar{Q}$ cohomology}\label{caq}

In this paper we discuss a slightly different type of algebra that is present in the model (\ref{J1}) for  generic $\m$. This is the chiral algebra generated by the cohomology classes of one of the supercharges.
Explicitly, the SYK model we are interested in have two supercharges 
\be
 Q =\frac{\partial}{\partial \th^+}- 2\bar{\th}^+\partial_z\,,\qquad
\bar{Q} =-\frac{\partial}{\partial \bar{\th}^+}+ 2{\th}^+\partial_z\,,
\ee
that satisfy
\be
\{Q ,\bar{Q} \}=4 \pa_z\,,\qquad Q^2=0\,,\qquad \bar{Q}^2=0\ .\label{qqb}
\ee
We are interested in the cohomology of one supercharge, say $\bar{Q} $.  
Following (\ref{qqb}), the elements of the cohomology can only have $\bar{z}$ dependence, up to $\bar{Q}$ exact terms. 
As a result, the elements of the cohomology generate a chiral algebra in the antiholomorphic sector \cite{Witten:1993jg,Witten:2005px}, see also~\cite{Dedushenko:2015opz}. It is this chiral algebra that we would like to study in the 2 dimensional supersymmetric SYK models. \footnote{Notice that we consider the $\bar{Q}$ cohomology, which is in a different notation from that used in \cite{Witten:1993jg,Witten:2005px}.}

Such chiral algebras have been discussed for Landau-Ginzburg models, see e.g.  \cite{Witten:1993jg,Kachru:1993pg,Kawai:1994np,Dedushenko:2015opz}. One crucial property of the chiral algebra, which follows from the fact that the algebra is defined in the cohomology of $\bar{Q} $, is that  details of the interaction do not affect the form of the chiral algebra \cite{Witten:1993jg,Kachru:1993pg}. In particular, the chiral algebra admits free field representations in terms of the $(\b,\gamma)$-ghost system for the chiral supermultiplet and the $(b,c)$-ghost system for the Fermi multiplet. This is also true for our model. A similar argument following \cite{Witten:1993jg,Kawai:1993jk,Kachru:1993pg,Kawai:1994np} indicates that the chiral algebra of our model can also be constructed in terms of the ghost fields.
In the following sections, we explicitly construct some of the generators of the chiral algebra with higher spin. We choose to use the set of 
\be
\phi^a,\quad \bar{\pa}\bar{\phi}^a,\quad  \lambda^i,\quad \bar{\lambda}^i\,,\label{freefields}
\ee
as the fundamental building blocks.\footnote{Our results can also be written in terms of the ghost fields  via the following dictionary 
\be\label{togh}
\phi^a \Leftrightarrow \gamma^a\,,\qquad \bar{\pa}\bar{\phi}^a\Leftrightarrow  \beta^a\,,\qquad \lambda^i\Leftrightarrow b^i\,,\qquad \bar{\lambda}^i\Leftrightarrow c^i\ .
\ee }
Furthermore, since the interaction term does not directly affect the form of the algebra, such algebras exist for each individual realization of the random coupling. Therefore our result does not rely on averaging over the random coupling, which makes our conclusion applicable to the full quantum mechanical model and avoids possible subtleties about the replica symmetries, see e.g.~\cite{Gur-Ari:2018okm,Ye:2018qzw,Arefeva:2018vfp,Wang:2018ijz}.

On the other hand, although the chiral algebra in the cohomology of $\bar{Q}$ does not directly depends on the form of the interaction \cite{Kachru:1993pg},  the form of the interaction (\ref{J1}) does determine the global symmetries the model has, which imposes nontrivial conditions on the chiral algebras of the model. 

Firstly, our model has a $U(1)_R$ symmetry that transforms the various fields according to
\be
\theta^+\to e^{i\ep }\theta^+\,, \qquad \Lambda^i\to e^{i\ep \tilde{\a}_i}\Lambda^i\,,\qquad  \Phi^a \to e^{i\ep {\a}_a}\Phi^a\ .
\ee
The fact that each term in the sum of the interaction is invariant under this symmetry means 
\be
\tilde{\a}_i+\sum_{k=1}^{q}\a_{a_k} =1\,, \qquad 1\leq  a_k\leq N\,,\qquad 1\leq  i\leq M\ .\label{Rcharge}
\ee
Since the interaction runs over all combination of the fermions, which means there is one equation (\ref{Rcharge}) for each set of indices $\{i,a_1,\ldots,a_q\}$. This large set of equations is only solved when
\be
\al_a=\al\,,\qquad \tilde\al_a=\tilde\al\,,
\ee
which satisfy
\be
\tilde{\a}+q\a =1\ .\label{RR}
\ee
Following almost identical argument as in~\cite{Witten:1993jg,Silverstein:1994ih}, this is the same condition to guarantee the existence of a left-moving stress tensor at the classical level.

In addition, there is another global $U(1)_L$ symmetry of the model
\bea
\Lambda^i\to e^{i\ep \tilde{p}_i}\Lambda\,,\quad  \Phi^a \to e^{i\ep {p}_a}\Phi^a\,, \quad  \bar{\Lambda}^i\to e^{-i\ep \tilde{p}_i}\bar{\Lambda}\,,\quad  \bar{\Phi}^a \to e^{-i\ep {p}_a}\bar{\Phi}^a\ .
\eea
The invariance of the action under this symmetry requires the charges to satisfy
\be
p_a=p\,,\qquad \tilde{p}_i=\tilde{p}\,,\qquad \tilde{p}+	q p =0\ . \label{RL}
\ee 
These relations among the charges give another set of nontrivial constraints on the form of the generators of the chiral algebra.

Secondly, the interaction (\ref{J1}) breaks the complete permutation symmetry, and hence any non-abelian global symmetry, among the $N$ chiral multiplets and the $M$  Fermi multiplets. Therefore the  operators in the chiral algebra do not need to respect such symmetries. This allows more general form of the chiral algebra, which is the crucial point so that a direct connection to the ``higher spin square" is possible.

\subsection{Enhance to $\mathcal{N}=(2,2)$ supersymmetry}

The model with $M=N$ has an enhanced $\mathcal{N}=(2,2)$ supersymmetry,  and the model reduces to that discussed in \cite{Murugan:2017eto,Bulycheva:2018qcp}. 

In this case the $U(1)_R$ and $U(1)_L$ symmetry presented in the $\mathcal{N}=(0,2)$ model combines to the $U(1)_v$ R-symmetry of the $\mathcal{N}=(2,2)$ model
\be
U(1)_v=U(1)_L\oplus U(1)_R\ .
\ee 
In particular, since the supersymmetry is enlarged, the chiral and Fermi multiplets are combined into $\mathcal{N}=(2,2)$ chiral supermultiplet, so their charges are related. In particular, we get
\be
\alpha=\tilde{\alpha}\,, \qquad p=\tilde{p}+1\,,\label{n221}
\ee
and the charge of the $\mathcal{N}=(2,2)$ supermultiplet under the $U(1)_v$ symmetry is simply
\be
r=\alpha+p\ . 
\ee
As a consistency check, at $M=N$ the condition 
(\ref{RR}) and (\ref{RL})
leads to
\be
\alpha=\frac{1}{q+1}\,,\qquad p=\frac{1}{q+1}\ .\label{n222}
\ee
The total R-charge under the $U(1)_v$ symmetry then becomes 
\be
r=\frac{2}{q+1}\,,
\ee
which is the same as the R-charge of the Landau-Ginzburg model with a $\Phi^{q+1}$ potential.

Apart from this, the rest computation of the chiral algebra at $M=N$ is largely parallel to those of the $\mathcal{N}=(0,2)$ model which we discuss in the next section.

\section{Minimal subalgebras with higher-spin extension} \label{walgcmp}

\subsection{The 
	$\mathcal{N}=(0,2)$ SYK model}\label{n02}

In this section we work out the first few operators of the chiral algebra and determine the OPE among them.

As discussed above, the higher-spin operators that generate the chiral algebra can be expressed in terms of  the free fields (\ref{togh}) 
of the chiral multiplet and the Fermi multiplet 
with the OPEs \footnote{In the following we use the Mathematica package developed by Thielemans \cite{Thielemans:1991uw} to perform some of the OPE computations.}
\bea
\bar{\phi}^a(x) \, \phi^b(y)   =  \delta^{ab} \,
\log [ (x-y)(\bar{x}-\bar{y})],
\label{fbf} && \quad 
\bar{\la}^i(x) \, \la^j(y)   =  \frac{-2}{(\bar{x}-\bar{y})}  \,
\delta^{ij}.
\label{fundOPE}
\eea

\subsubsection*{$\bullet$ Spin-2}
The (antiholomorphic) stress energy tensor of the model~(\ref{J1}) is
\be
T(\bar{z}) = \sum_{a=1}^N \Bigg[ (1-\frac{\al_a}{2}) \, \bar{\pa} \phi^a \, \bar{\pa}
\bar{\phi}^a -\frac{\al_a}{2} \, \phi^a \, \bar{\pa}^2 \bar{\phi}^a
\Bigg](\bar{z}) +\sum_{i=1}^M \Bigg[ \frac{1+\tilde{\al}_i}{4}  \, \la^i \, \bar{\pa}
\bar{\la}^i -\frac{1-\tilde{\al}_i }{4} \, \bar{\pa} \la^i \, \bar{\la}^i
\Bigg](\bar{z})\,,
\label{T}
\ee
with the standard OPE 
\bea
T(\bar{z}) \, T(\bar{w}) \sim \frac{1}{(\bar{z}-\bar{w})^4} \, \frac{c}{2} +
\frac{1}{(\bar{z}-\bar{w})^2} \, 2 T(\bar{w}) + \frac{1}{(\bar{z}-\bar{w})} \, \bar{\pa} \, T(\bar{w})
,
\nonu
\eea
where
the central charge is given by
\bea
c  & = & \sum_{a=1}^{N} (2-6\al_a + 3\al_a^2) +\sum_{i=1}^{M} (1 - 3
\tilde{\al}_i^2).
\label{c}
\eea
With respect to this stress energy tensor the conformal dimensions for $\phi^a(z)$, $\bar{\pa} \, \bar{\phi}^a(z)$, $\la^i(z)$
and $\bar{\la}^i(z)$ are 
$\frac{1}{2} \al_a$, $\frac{1}{2}(2-\al_a)$, $\frac{1}{2}(1+\tilde{\al}_i)$
and $\frac{1}{2}(1-\tilde{\al}_i)$ respectively.
The $\al_a$ and $\tilde{\a}_a$ parameters take the values in~(\ref{RR}) and (\ref{RL}). In particular, the central charge is
\bea
c  & = & N (2-6\al + 3\al^2) +M (1 - 3
\tilde{\al}^2)\ . 
\label{cSYK}
\eea
Notice that our stress energy tensor takes a similar form as those of the Landau-Ginzburg model \cite{Silverstein:1994ih,Dedushenko:2015opz}. 

\subsubsection*{$\bullet$ Spin-1}
There is a spin-1  operator corresponding to the $U(1)_L$ symmetry 
\bea
J(\bar{z}) =  \sum_{a=1}^N \, p_a \, \phi^a \, \bar{\pa} \bar{\phi}^a(\bar{z}) -\sum_{i=1}^M \, \frac{\tilde{p}_i}{2} \, \la^i \, \bar{\la}^i(\bar{z})\,,
 \label{J}
\eea
which is a Virasoro primary operator
of spin $1$ under the stress
energy tensor
\bea
T(\bar{z}) \, J(\bar{w}) \sim
\frac{1}{(\bar{z}-\bar{w})^2} \,  J(\bar{w}) + \frac{1}{(\bar{z}-\bar{w})} \, \bar{\pa} \, J(\bar{w})\,,
\label{tj}
\eea
if the following condition is satisfied
\bea
\sum_{a=1}^{N} \, p_a \, (1 -\al_a) +\sum_{i=1}^{M} \, \tilde{p}_i
\tilde{\al}_i =0.
\label{polethree}
\eea
This together with~(\ref{RL}) fixes 
$\tilde{\a}$ parameter to
\be
\tilde{\a}=\frac{N (q-1)}{M q^2-N}\ .
\ee 
In the following, we focus on the model with this value of the parameter. In particular, the central charge for this case becomes
\be
c= \frac{N^2 + M^2 q^2 + 2 M N (1 - 3 q + q^2)}{-N + M q^2}\ .
\ee
The level of the $U(1)$ current algebra is determined by the  OPE 
\bea
J(\bar{z}) \, J(\bar{w}) \sim\frac{c_{J}}{(\bar{z}-\bar{w})^2} \, ,\qquad 
c_J = \sum_{i=1}^{M} \, \tilde{p}_i^2-\sum_{a=1}^{N} \, p_a^2.
\label{cj}
\eea
The $U(1)$ charges for
$\phi^a(z)$, $\bar{\pa} \, \bar{\phi}^a(z)$, $\la^i(z)$
and $\bar{\la}^i(z)$ are 
$ p_a$, $- p_a$, $ \tilde{p}_i$, and $- \tilde{p}_i$
respectively.
For our SYK model, the charges obey  (\ref{RR}) and (\ref{RL}). This means
\be
c_J = M \, \tilde{p}^2 - N \, p^2\,,
\ee
and the condition (\ref{polethree}) reduces to 
\bea
{N} \, p \, (1 -\al) +{M} \, \tilde{p}
\tilde{\al}  =0.
\label{polethreeSYK}
\eea

The $U(1)_R$ current is in the same supermultiplet with the stress energy tensor and does not lead to new operator in the cohomological algebra. There is no other conserved spin-1 generator that leads to a symmetry of the model.

\subsubsection*{$\bullet$ Spin-3}

Consider the following ansatz for a spin-3 operator
\bea
W_3(\bar{z})  & = &
c_1   \, \bar{\pa}^2 \, \phi^a \, \bar{\pa} \bar{\phi}^a(\bar{z})
+c_2   \, \bar{\pa} \, \phi^a \, \bar{\pa}^2 \, \bar{\phi}^a(\bar{z})
+c_3   \, \phi^a \, \bar{\pa}^3 \, \bar{\phi}^a(\bar{z})+c_{14} \,  
{\la}^i   \, \bar{\pa}^2 \, \bar{\la}^i(\bar{z})
\nonu \\
&+& c_5   \, \phi^a \, \bar{\pa} \, \bar{\phi}^a
\, \phi^b \, \bar{\pa}^2\, \bar{\phi}^b(\bar{z})
+  c_6   \, \phi^a \, \bar{\pa} \, \bar{\phi}^a
\, \phi^b \, \bar{\pa} \, \bar{\phi}^b \,
\phi^c \, \bar{\pa} \, \bar{\phi}^c(\bar{z})
\nonu \\
& + &  c_7    \,
\phi^a \, \bar{\pa} \, \bar{\phi}^a
\, \phi^b \, \bar{\pa} \, \bar{\phi}^b \,
\la^i  \, \bar{\la}^i(\bar{z})
+   c_8    \,
\phi^a \, \bar{\pa} \, \bar{\phi}^a \,
\bar{\pa} \, \bar{\la}^i  \, \la^i(\bar{z})
\nonu \\
&+& c_9    \,
\phi^a \, \bar{\pa} \, \bar{\phi}^a \,
\bar{\la}^i  \, \bar{\pa} \, \la^i(\bar{z})
+ c_{10}   \,
\bar{\pa} \, \phi^a \, \bar{\pa} \, \bar{\phi}^a \,
\bar{\la}^i   \, \la^i(\bar{z})
\nonu \\
&+ &  c_{11}   \,
\phi^a \, \bar{\pa}^2 \, \bar{\phi}^a \,
\bar{\la}^i   \, \la^i(\bar{z})
+  c_{12} \,  
\bar{\pa}^2 \, {\la}^i   \, \bar{\la}^i(\bar{z})
+  c_{13} \,   
\bar{\pa} \, {\la}^i   \, \bar{\pa}\, \bar{\la}^i(\bar{z})
\nonu \\
& + &   c_{15} \,  
{\la}^i   \, \bar{\la}^i \,
{\la}^j   \, \bar{\la}^j \,
{\la}^k   \, \bar{\la}^k(\bar{z}) 
+  c_{16}   \,
\phi^a \, \bar{\pa} \, \bar{\phi}^a \,
{\la}^i   \, \bar{\la}^i
\,  {\la}^j   \, \bar{\la}^j(\bar{z})
\nonu \\
& + & c_4 \,   \, \phi^a \, \bar{\pa} \, \bar{\phi}^a
\, \bar{\pa} \, \phi^b \, \bar{\pa} \, \bar{\phi}^b(\bar{z})
+  c_{17} \,    \,
{\la}^i   \, \bar{\la}^i
\,  \bar{\pa} \, {\la}^j   \, \bar{\la}^j(\bar{z})
+   c_{18} \,    \,
{\la}^i    \, \bar{\la}^i
\,  {\la}^j   \, \bar{\pa} \, \bar{\la}^j(\bar{z})\,,
\label{W3}
\eea
where repeated indices are summed over.
We look for a spin-$3$ Virasoro primary operator with
\bea
T(\bar{z}) \, W_3(\bar{z}) \sim
\frac{1}{(\bar{z}-\bar{w})^2} \, 3 \, W_3(\bar{z}) + \frac{1}{(\bar{z}-\bar{w})} \, \bar{\pa} \, W_3(\bar{z}) \ .
\label{tw3}
\eea
This fixes some of the coefficients in the ansatz (\ref{W3}). Further requiring  its OPE with the spin-1  operator to be regular, as well as its OPE with itself closes in the algebra, we obtain further conditions that fix all but three coefficients. One of the remaining coefficient corresponds to the normalization of the spin-3 operator and can be fixed by 
\be
W_3(\bar{z})W_3(\bar{w})\sim \frac{c_{W_3}}{(\bar{z}-\bar{w})^6}+\mathcal{O}((\bar{z}-\bar{w})^{-5})\ .
\ee
In the following we choose the canonical normalization, namely $c_{W_3}=(c-1)/3$, after the spin-1 operator is factored out. 
One can then read off the spin-4 operator from the second order pole of the $W_3(\bar{z})W_3(\bar{w})$ OPE. Further requiring the spin-4 operator to be Virasoro primary, together with the requirement of its OPE with the spin-1 is regular, fixes all but one parameter in the ansatz~(\ref{W3}). 
After applying all these conditions, the $W_3(z) \, W_3(w)$ OPE reads
\bea
 W_3 (\bar{z}) \, W_3(\bar{w})  & \sim&  \frac{c_{W_3}}{(\bar{z}-\bar{w})^6} \, +
\frac{\frac{6 \, c_{W_3}}{(c-1)} \, \tilde{T}(\bar{w})}{(\bar{z}-\bar{w})^4} +  \frac{ \frac{3 \, c_{W_3}}{(c-1)} \, \bar{\pa} \, \tilde{T}(\bar{w})}{(\bar{z}-\bar{w})^3}  +
\frac{	\frac{ 9\, c_{W_3}}{10(c-1)} \, \bar{\pa}^2 \, \tilde{T}}{(\bar{z}-\bar{w})^2} +
\frac{	\frac{ \, c_{W_3}}{5(c-1)} \, \bar{\pa}^3 \, \tilde{T}}{\bar{z}-\bar{w}} \,\qquad  
\nonu \\
&& +  \frac{1}{(\bar{z}-\bar{w})^2} \frac{44 \, c_{W_3}}{(c-1)\,(5c+17)} \,
\left(( \tilde{T} \, \tilde{T} -\frac{3}{10} \, \bar{\pa}^2 \tilde{T})(\bar{w}) 
+ {W}_4(\bar{w})\right)
\nonu \\
&& +
\frac{1}{\bar{z}-\bar{w}} \,  \frac{22 \, c_{W_3}}{(c-1)\,(5c+17)} \,
 \left(\bar{\pa} \,( \tilde{T} \, \tilde{T} -\frac{3}{10} \, \bar{\pa}^2 \tilde{T})(\bar{w}) 
+ \bar{\pa} \,{W}_4(\bar{w})\right)\,,
 \label{w3w3}
\eea
where a new spin-$4$ operator $W_4(\bar{w})$ appears. In the above expression we have also defined 
\be
\tilde{T}(\bar{z}) = T(\bar{z}) - \frac{1}{2c_J} (J J)(\bar{z})\,,
\label{newT}
\ee
that have regular OPE with the spin-1 operator $J(\bar{z})$. 
The explicit form of the $W_4$ operator is not quite illuminating, so we omit it for simplicity. Notice that this spin-4 operator still depends on the remaining undetermined parameter. The above $W_3(\bar{z}) \, W_3(\bar{w})$  OPE agrees with the known result in the literature, say the one in \cite{Bais:1987zk} once we identify the $\tilde{c}$ there as $c-1$ in our expressions.

One might then propose that we can continue to construct operators with higher spins and to find a $\mathcal{W}_{\infty}$ algebra. However, we claim that the algebra generated in this way is not a conventional $\mathcal{W}_{\infty}$ algebra.  The reason is the following. 
The conditions mentioned above are all that we could impose to get a closed algebra 
among the first few low spin operators. However, the fact that we cannot fix one parameter in the ansatz~(\ref{W3}) is quite surprising. Indeed, the unfixed parameter in the spin-3 ansatz signals the existence of two different spin-3 primary operators. The parameter also appears in the spin-4 operator, which indicates that there are at least two spin-4 operators in the algebra. On the other hand, we know from the result in~\cite{Gaberdiel:2012ku} that the family of bosonic $\mathcal{W}_{\infty}[\lambda]$ algebra is described by only two parameters, which relies on the assumption that the set of operators that generate the higher-spin algebra contains only one higher-spin operator for each given spin. Therefore the fact that there are more than one operators at each large enough spin is consistent with the fact that there are more parameters than the conventional $\mathcal{W}_\infty$ algebra. This implies that what we have found is not a higher-spin type algebra. In other words, the minimal subalgebra of this chiral algebra that contains at least one higher-spin primary operator is larger than the conventional higher-spin  algebra. 

We believe the above statement is correct. A simple test of this claim would be to compute higher order OPEs, such as the $W_3(\bar{z})W_4(\bar{w})$ and the $W_4(\bar{z})W_4(\bar{w})$ OPE, and test if the parameter does appear explicitly in  these OPEs.  Unfortunately, we do not have enough computational power to confirm this. We hope to come back to this computation in the future with an updated power of computation.

Nevertheless, we can test our statement indirectly by considering a special model at $M=N$, where the model has an $\mathcal{N}=(2,2)$ supersymmetry.  In that case due to the larger symmetry, the result is more constrained and we can  observe the existence of extra operators, comparing to the conventional $\mathcal{SW}_{\infty}$ algebra in e.g.~\cite{Bergshoeff:1990cz},  explicitly. Since the $\mathcal{N}=(2,2)$ is a special case of the $\mathcal{N}=(0,2)$ model, the result in the next section provides a test of the prediction we made on a special case.

\subsection{The 
	$\mathcal{N}=(2,2)$ SYK model }\label{n22}

A special case of the model (\ref{J1}), namely at $M=N$, has an enhanced $\mathcal{N}=(2,2)$ supersymmetry. This enhancement renders the cohomological chiral algebra supersymmetric. 
In the following we construct the first few higher-spin operators of this model explicitly. 

Due to the presence of supersymmetry, the operators are organized into supermultiplets. The lowest supermultiplet contains a spin-1 current operator, two supercharges and the stress energy tensor. This multiplet of the $\mathcal{N}=(2,2)$ Landau-Ginzburg model is worked out in e.g. \cite{Witten:1993jg}. The operators in our model are similar, the only difference is that the charges   
$\al_a$, $\tilde{\al}_i$, $p_a$ and $\tilde{p}_i$ take the value in (\ref{n221}) and (\ref{n222}). Explicitly, the operators in this multiplet of our SYK model~(\ref{J1}) are
\bea
J(\bar{z}) &=& \frac{ q}{2(q+1)}\sum_{i=1}^N \,   \bar\la^i \, {\la}^i(\bar{z})
-\frac{1}{q+1} \sum_{a=1}^N   \, \phi^a \, \bar{\pa} \bar{\phi}^a(\bar{z})\,, \\
G^{+}(\bar{z}) & = & \sum_{a=1}^{N} \, \frac{1}{\sqrt{2}}
\, \bar{\pa} \, \bar{\phi}^a \, \la^a(\bar{z})\,,
\\
G^{-}(\bar{z}) & = & -\frac{\sqrt{2} q}{(1+q)}
\, \Bigg[ \sum_{a=1}^{N} \, \bar{\pa} \, \phi^a \,
\bar{\la}^a -\frac{1}{q} \, \sum_{a=1}^{N} \, \phi^a \, \bar{\pa} \,
\bar{\la}^a \Bigg](\bar{z})\,,
\label{Gpm}\\
T(\bar{z}) &=& \frac{1}{(2q+2)}
\sum_{a=1}^N \Bigg[ {(2q+1)} \, \bar{\pa} \phi^a \, \bar{\pa}
\bar{\phi}^a -\, \phi^a \, \bar{\pa}^2 \bar{\phi}^a
- \frac{(q+2)}{2}   \, \bar{\pa}
\bar{\la}^a\, \la^a +\frac{q}{2}   \, \bar{\la}^a\, \bar{\pa} \la^a
\Bigg](\bar{z}).\qquad \quad \label{T22}
\eea
The central charge is given by
\bea
c = \frac{3N (q-1)}{(q+1)} \ .\label{cc}
\eea
The generators $ J(\bar{z})$,
$G^{\pm}(\bar{z})$, and $T(\bar{z})$ satisfy  
the standard ${\cal N}=2$ superconformal algebra.

Next we consider the higher-spin operators.
The first higher-spin $\mathcal{N}=2$ multiplet consists of operators with spin
$(2,\frac{5}{2}, \frac{5}{2},3)$ . We only need to determine the new spin-2 operator and the other operators are its $\mathcal{N}=2$ supersymmetric descendants. The superconformal primary spin-2 operator reads
\bea
\nonumber  W_2(\bar{z})  &=& (2 + q) ( 2 q-1) \,  \bar{\pa} \phi^a \, \bar{\pa} \bar{\phi}^a(\bar{z})
+ (2 - q) \,   \phi^a \, \bar{\pa}^2 \bar{\phi}^a(\bar{z})
+ \frac{1}{2} (1-2 q) q   \, \la^i \, \bar{\pa} \bar{\la}^i(\bar{z})\qquad \quad\\
\nonu && ~ + \frac{1}{2} (q-2) (2 q+1)  \,
\bar{\pa} \la^i \,  \bar{\la}^i(\bar{z}) + \frac{2 (2-q) q}{(N-1) (q-1)} \,  \phi^a \,
\bar{\pa} \bar{\phi}^a \,
\,  \bar{\la}^i\la^i (\bar{z})\\
\nonumber&&~ + \frac{2 (q-2)}{(N-1) (q-1)} \,    \, \phi^a \, \bar{\pa} \bar{\phi}^a \,
\phi^b \, \bar{\pa} \bar{\phi}^b(\bar{z}) +
\frac{(q-2) q^2}{2 (N-1) (q-1)} \,  \la^i \,
\bar{\la}^i \, \la^j \, \bar{\la}^j(\bar{z})\\
&&~ +\frac{(q+1-3 N (q-1)) }{(N-1) (q-1)}   \, \left(\phi^a\phi^a \bar{\pa}\bar{\phi }^a\bar{\pa}\bar{\phi }^a(\bar{z})-q\phi^a\bar{\pa}\bar{\phi }^a\bar{\lambda }^i\lambda^i(\bar{z})\right)\ .
\label{W2twotwo}
\eea
We expect the inclusion of this multiplet leads to an infinite dimensional algebra~\cite{Bergshoeff:1990yd}.  
We can proceed to compute the $W_2(\bar{z})W_2(\bar{w})$ OPE
\bea
W_2(\bar{z})W_2(\bar{w})&\sim& \frac{n_2}{(\bar{z}-\bar{w})^4} +\frac{\frac{4n_2}{c-1}T(\bar{w})-\frac{6n_2}{c(c-1)}(JJ)(\bar{w})+c_{22,2} W_2(\bar{w})}{(\bar{z}-\bar{w})^2}\\
&&+\frac{\frac{2n_2}{c-1}\bar{\pa} T(\bar{w})-\frac{6n_2}{c(c-1)}(J\bar{\pa} J)(\bar{w})+\frac{c_{22,2}}{2} \bar{\pa} W_2(\bar{w})}{(\bar{z}-\bar{w})}\,,
\eea
where the central charge is~(\ref{cc}) and
\bea
n_2&=&\frac{4 N (q-2) (q+1)^2 (3 N (q-1)-q-1)}{N-1}\,,\\
c_{22,2}&=&-\frac{4 (q+1) (N (q-5)+q+1)}{N-1}\ .
\eea
The form of this OPE is identical to equation~(3.30) of~\cite{Candu:2012tr}, so 
one is tempting to identify a quartic relation between the $\lambda$ 
parameter of the $\mathcal{SW}_{\infty}[\lambda]$ algebra and the $q, N$ 
parameter in our model
\bea
\nonumber \lambda ^4+\frac{\lambda ^3 (q-1) (N (q-1)+q+1)^2}{2 (q+1) (N (q-3)+q+1)}+\frac{N^2 (q-1)^2}{(q+1)^2}-\frac{\lambda  N (q-1)^2 (N (q-1)+q+1)^2}{2 (q+1)^2 (N (q-3)+q+1)}\qquad \qquad  &&\\
\nonumber  +\frac{\lambda ^2 (q-1) \left(N^3 (q-1)^3+N^2 (q+1) (q (q+2)-11)-N (q-5) (q+1)^2-(q+1)^3\right)}{2 (q+1)^2 (N (q-3)+q+1)}=0\ .\quad && 
\eea  
But as we will argue in the following this mapping might not be very meaningful.

The problem is the following. One can work out the other components of the 
multiplet, for example the two fermionic components in the multiplet are
\bea
\nonumber W_{5/2}^+(\bar{z})&=&\sqrt{2} (q+1) (2 q-1)\Bigg(\frac{2 (q-2) \phi ^b\bar{\pa} \bar{\phi }^a\bar{\pa}\bar{\phi }^b\lambda^a}{(N-1) (q-1) (2 q-1)}-\frac{(q-2) q \bar{\pa}\bar{\phi }^b\bar{\lambda }^a\lambda^a\lambda^b}{(N-1) (q-1) (2 q-1)}\qquad \\
&&-\frac{(3 N q-3 N-q-1) \phi ^a\bar{\pa}\bar{\phi }^a\bar{\pa}\bar{\phi }^a\lambda^a}{(N-1) (q-1) (2 q-1)}+\left(\bar{\pa}\bar{\phi }^i\bar{\pa}\lambda ^i\right)-\frac{(q-2) \bar{\pa}^2\bar{\phi }^i\lambda^i}{2 q-1}\Bigg)(\bar{z})\,,\\
\nonumber W_{5/2}^-(\bar{z})&=&\sqrt{2} (q-2)\Bigg(\frac{2 q^2 \bar{\pa}\phi^j\bar{\lambda }^i\bar{\lambda}^j\lambda^i}{(N-1) (q-1)}+\frac{4 q \bar{\pa}\phi ^i\phi ^j\bar{\pa}\bar{\phi }^j\bar{\lambda }^i}{(N-1) (q-1)}-\frac{2 q \phi ^j\bar{\lambda }^i\bar{\pa} \bar{\lambda }_j\lambda ^i}{(N-1) (q-1)}\\
\nonumber &&-\frac{4 \phi ^i\phi ^j \bar{\pa}\bar{\phi }^i\bar{\pa}\bar{\lambda }^j}{(N-1) 
(q-1)}-\frac{q (3 N q-3 N-q-1) \phi ^i\bar{\pa}\bar{\lambda }^i\bar{\lambda 
}^i\lambda ^i}{(N-1) (q-2) (q-1)}\\
\nonumber &&-\frac{2 q (3 N q-3 N-q-1) \bar{\pa}\phi ^i\phi ^i\bar{\pa}\bar{\phi }^i\bar{\lambda }^i}{(N-1) (q-2) (q-1)}+\frac{2 (3 N q-3 N-q-1) \phi ^i\phi ^i\bar{\pa}\bar{\phi }^i\bar{\pa}\bar{\lambda }^i}{(N-1) (q-2) (q-1)}\\
&&+\phi ^i\bar{\pa}^2\bar{\lambda }^i-\frac{4 (q-1) (q+1) \bar{\pa}\phi ^i\bar{\pa}\bar{\lambda }^i}{q-2}+\frac{(2 q-1) q \bar{\pa}^2\phi ^i\bar{\lambda }^i}{q-2}\Bigg)(\bar{z})\,,
\eea
where repeated indices are all summed over.
One can then compute the $W^+_{5/2}(\bar{z})W^+_{5/2}(\bar{w})$ OPE. If the 
underlying algebra is indeed a $\mathcal{SW}_{\infty}[\lambda]$ algebra~\cite{Bergshoeff:1990cz,Bergshoeff:1990yd}, one expects 
that the right-hand-side of this OPE should only contain normal ordered products of operators with smaller dimension since there is no generator in the $\mathcal{SW}_{\infty}[\lambda]$ algebra with two units of the spin-1 charge. 

Surprisingly, this is not the case for our algebra. One can compute the  $W^+_{5/2}(\bar{z})W^+_{5/2}(\bar{w})$ OPE explicitly and get
\bea
W^+_{5/2}(\bar{z})W^+_{5/2}(\bar{w})&\sim& \frac{\frac{16 (q-2) (q+1)^4}{(N-1) (q-1)}(G^+\bar{\pa} G^+)(\bar{w})-\frac{8 (q+1)^2 (N (q-5)+q+1)}{(N-1) (N (q-1)+q+1)}(G^+ W_{5/2}^+)(\bar{w})}{\bar{z}-\bar{w}}\quad \nonumber\\
&&\quad -\frac{\frac{4 (q+1)^3 (-3 N (q-1)+q+1)^2 }{(N-1)^2 (q-1) (N (q-1)+q+1)}  W_4^{++}(\bar{w})}{\bar{z}-\bar{w}}\,,
\eea
where the new Virasoro primary spin-4 operator
\be
W_4^{++}(\bar{z})=2 \left(\bar{\pa}\bar{\phi }^i \bar{\pa}\bar{\phi }^j \lambda^i\bar{\pa}\lambda^j\right)(\bar{z})+(N+1) \left(\bar{\pa}\bar{\phi }^i \bar{\pa}\bar{\phi }^i \bar{\pa}\lambda^i \lambda^i\right)(\bar{z})+\frac{2 }{q-1}\left(\phi ^i\bar{\pa}\bar{\phi }^i\bar{\pa}\bar{\phi }^i \bar{\pa}\bar{\phi }^j \lambda^i\lambda^j\right)(\bar{z})\,,\qquad
\ee
carries two units of the spin-1 charges. One can further check that it is a descendant of a spin-$7/2$ supersymmetric primary operator with one unit of spin-1 charge
\bea
\nonumber W_{\frac{7}{2}}^+(\bar{z})&=&2(q+2) \left(\bar{\pa}\phi ^j\bar{\pa}\bar{\phi }^i\bar{\pa}\bar{\phi }^j\lambda ^i\right)(\bar{z})+q\left(\bar{\pa}\bar{\phi }^j\bar{\lambda }^i\lambda ^i\bar{\pa}\lambda ^j\right)(\bar{z})+\frac{q }{q-1}\left(\phi ^i\bar{\pa}\bar{\phi }^i\bar{\pa}\bar{\phi }^i\bar{\lambda }^j\lambda ^i\lambda ^j\right)(\bar{z})\\
\nonumber &&+\frac{2 q }{q-1}\left(\phi ^i\bar{\pa}\bar{\phi }^i\bar{\pa}\bar{\phi }^j\bar{\lambda }^i\lambda ^i\lambda ^j\right)(\bar{z})-2\left(\phi ^j\bar{\pa}\bar{\phi }^i\bar{\pa}\bar{\phi }^j\bar{\pa}\lambda ^i\right)(\bar{z})+(N+1) q  \left(\bar{\pa}\bar{\phi }^i\bar{\lambda }^i\bar{\pa}\lambda ^i\lambda ^i\right)(\bar{z})\\
\nonumber &&-\frac{2  }{q-1}\left(\phi ^i\phi ^i\bar{\pa}\bar{\phi }^i\bar{\pa}\bar{\phi }^i\bar{\pa}\bar{\phi }^j\lambda ^j\right)(\bar{z})+(N+1  )\left(\phi ^i\bar{\pa}\bar{\phi }^i\bar{\pa}\bar{\phi }^i\bar{\pa}\lambda ^i\right)(\bar{z})-q \left(\bar{\pa}\bar{\phi }^j\bar{\lambda }^i\bar{\pa}\lambda ^i\lambda ^j\right)(\bar{z})\\
&&-(N+1)  (q+2) \left(\bar{\pa}\phi ^i\bar{\pa}\bar{\phi }^i\bar{\pa}\bar{\phi }^i\lambda ^i\right)(\bar{z})+\frac{2  }{q-1}\left(\phi ^i\phi ^j\bar{\pa}\bar{\phi }^i\bar{\pa}\bar{\phi }^i\bar{\pa}\bar{\phi }^j\lambda ^i\right)(\bar{z})\ .
\eea
It is easy to check that they belong to a multiplet with spin content $(\frac{7}{2},4,4,\frac{9}{2})$. Therefore, we explicitly observe that the subalgebra we identified, which is the minimal one that contains a higher-spin operator, contains more operators than a conventional higher-spin $\mathcal{SW}_{\infty}$ algebra.    
In addition, based on the pattern in which unexpected multplets appear, the number of supersymmetric multiplets seems to grow as spin increases.

 \subsection{Anomaly }
 
 As discussed in section~\ref{caq}, at the classical level, the existence of a stress energy 
 tensor in the algebra of $\bar{Q}$ cohomology is guaranteed by the 
 quasi-homogeneity condition of the potential $\Lambda^i J_i(\Phi^a)$. Now that we have constructed a few operators in the chiral algebra, we can test if the result is anomaly free at the quantum level. We do not know a set of sufficient conditions for this, but we have a necessary condition to check following~\cite{Silverstein:1994ih}. We will take the example of the spin-2 Virasoro operator, and the computation for the other operators is similar.  
 At the quantum level, we would like to check if $[\bar{Q},T]=0$ is consistent 
 as an quantum operator identity. This amounts to check, for any operator 
 $O(x)$, the following identity 
 \be
 0=\int d^2 z \left(\partial_\mu S^\mu(z) T(\bar{w}) O(x)\right)\ .\label{cond1}
 \ee
 where $S^\mu$ is the supersymmetry current whose corresponding charge is 
 $\bar{Q}$ and  the appropriate ordering of the operators is implicitly 
 adopted. Following \cite{Silverstein:1994ih}, we can choose special operators 
 $O(x)$ where necessary condition for the quantum cohomology requirement could 
 be obtained from (\ref{cond1}). Explicitly, one rewrites (\ref{cond1}) into 
 the relation
 \be
 0=\left([\bar{Q},T(\bar{w})]O(x)\right)+\left(T(\bar{w})[\bar{Q},O(x)\}\right)\,,\label{sum}
 \ee
 where in the last term the parenthesis $[ \cdot \}$ denotes either commutator 
 or anticommutator depending on whether the  operator $O(x)$ is bosonic or 
 fermionic.
 To further evaluate this expression, we consider the super-derivatives 
 \be
 D=\frac{\partial}{\partial \th^+}+2\bar{\th}^+\partial_z\,,\qquad
 \bar{D}=-\frac{\partial}{\partial \bar{\th}^+}- 2{\th}^+\partial_z\,,
 \ee
 that are conjugates to the supercharges, namely
 \be
 Q=e^{4\th\bar\th \partial_z} D e^{-4 \th \bar\th\partial_z}\,,\qquad 
 \bar{Q}=e^{-4\th\bar\th \partial_z} \bar{D} e^{4 \th \bar\th\partial_z}\ . 
 \ee
 Therefore the $\bar{Q}$ cohomology is the same as the $\bar{D}$ cohomology 
 that is simpler to compute.
 Furthermore, as argued in \cite{Silverstein:1994ih}, one can use the the free 
 fields commutators,
 and the equation of motion
 \bea
 \bar{D} \pa_z \bar{\Phi}^a(z)=\Lambda^i \frac{\partial J_i(\Phi^a)}{\partial 
 \Phi^a}\,,\qquad \bar{D} \bar{\Lambda}^i = 2 J_i(\Phi^a)\,,\label{eom}
 \eea
 to compute the leading contribution to the OPEs.  
 
 The logic of the following computation is to choose some $O(x)$ operator whose 
 supersymmetry action $[\bar{Q},O(x)\}$ is known, then by computing the leading 
 order term of the second term of (\ref{cond1}) we get the value of the first  term
 \be
 \left([\bar{Q},T(\bar{w})]O(x)\right)\ .
 \ee
 We then check if the value of this term  is consistent with $[\bar{Q},T(\bar{w})]=0$.
 
 One simple choice is $O(x)=\lambda^i$. Then following the equation of motion 
 (\ref{eom}) the supersymmetry transformation of $O(x)$ is simply
 \be
 \{\bar{Q},\bar{\Lambda}^i\}=  2 J_i(\Phi^a)\ .
 \ee
 This gives
 \be\label{toev1}
 \left(T(\bar{w})\{\bar{Q},\lambda^i(x)\}\right)= 2 T(\bar{w})  J_i(\phi^a)(x) 
 \ .
 \ee
 Before doing the remaining OPE we first consider what would be the form 
 of the result to the leading correction of the coupling in the potential. This is all we need for the purpose of our 
 computation. 
 Since the sum of the two terms of~(\ref{sum}) must be zero, the leading term 
 should be canceled with the leading terms of the first term
 \be
 \left([\bar{Q},T(\bar{w})]\lambda^i(x)\right)\ .
 \ee
 The leading term of this expression only receives contributions from the free field 
 part, so the only singular term in this expression must come from the OPE 
 between $\lambda^i(x)$ and (the  derivatives of) any linear $\bar\lambda^i(w)$ 
 term. The result will be simply a constant term due to (\ref{fundOPE}). 
 Therefore a necessary condition for the vanishing of the anomaly in the stress 
 energy tensor is simply that the residual of the leading order pole of 
 (\ref{toev1}) is not proportional to the identity operator. 
 It is easy to check that this is true for our potential $J_i(\phi^a)$ since it 
 is holomorphic in the chiral superfields while the relevant terms in $T(\bar{w})$ contains both $\bar{\pa}\bar{\phi}^a$ and $\phi^a$.
 
 We can further consider the operator $O(x)=\phi^a$. Then it is straightforward 
 to check that it does not lead to any new condition since the result 
 $\{\bar{Q},\phi^a\}$ is Grassmann odd and thus its further OPE with 
 $T(\bar{w})$ will not be proportional to the identity operator.
 
 To sum up, we have shown that for the model (\ref{J1}) there could be an 
 anomaly free stress energy tensor at the quantum level. Similar arguments apply to the other higher-spin operators. Therefore we do not find 
 extra consistency conditions at quantum level.

\section{Higher spin square at the $q\to \frac{N}{M}$  limit}

A property of the higher-spin algebra defined in the previous section is that it always involves multiple-sum terms that are not normal ordered products of low spin operators. The inclusion of these multiple-sum terms is required by the closure of the algebra, which, however, makes the holographic interpretation of the higher-spin operators less clear.

On the other hand, as shown in~\cite{Peng:2018zap} explicitly,  there is good evidence showing that at a special value, $q\to \frac{N}{M}$, one observes the emergence of a tower of conserved operators in the infrared SYK-like fix point of the model~(\ref{J1}).  They correspond to a tower of massless higher-spin fields in the bulk which resist the early time chaotic behavior and make the Lyapunov exponent vanish. Given the property that the chiral algebra is rigid along the RG flow, one naturally expects to see a higher-spin algebra of single-sum operators which are holographically dual to the tower of ``single particle" higher-spin fields, to emerge in the chiral algebra. In addition, from the analysis of~\cite{Peng:2018zap} one expects this to appear only at the above special limit. 

So we search for higher-spin subalgebras of the chiral algebra that are generated by single-sum operators, which are analogues to the single-trace operators in matrix theories.   
It turns out that one can indeed find higher-spin ``single-sum" subalgebras of the chiral algebra, and these subalgebras only close at the special value $q\to\frac{N}{M}$. 
Moreover, we find two, instead of one, higher-spin subalgebras. They do not commute with each other; commutators of operators from the two different subalgebra  generate new operators. All the operators generated in such a manner can be organized into representations of the vertical higher-spin algebra. These properties indicate that at the special point $q\to\frac{N}{M}$ there is a  ``higher-spin square" structure that was previously found in the symmetric orbifold theory~\cite{Gaberdiel:2014cha, Gaberdiel:2015mra}. As expected there, see also a related example~\cite{Gaberdiel:2015uca}, the higher-spin square is a stringy algebra that is closely related to the tensionless limit of the string theory. Given our expectation of the relation between the SYK-like models and some finite tension string theory, and our interpretation of the limit $q\to\frac{N}{M}$ as a toy version of the tensionless limit, it is not surprising that we find such a ``higher-spin square" structure in this limit. 

In the following, we first discuss the two different higher-spin algebras. Then we discuss the higher-spin square structure they generate. At the end of this section, we comment on the relation between this algebra and the emergent higher-spin algebra discovered in~\cite{Peng:2018zap}.

\subsection{The ``vertical" higher-spin subalgebra} \label{hsv}
One of the higher-spin algebra at the limit $q\to\frac{N}{M}$ consists of operators that are quadratic in the fundamental fields. We call the subalgebra generated by them the ``vertical" higher-spin algebra in the notion of \cite{Gaberdiel:2014cha, Gaberdiel:2015mra}.

The stress energy tensor is the same as~(\ref{T}), which we recast here
\be
T(\bar{z}) = \sum_{a=1}^N \Bigg[ (1-\frac{\al_a}{2}) \, \bar\pa \phi^a \, \bar\pa
\bar{\phi}^a -\frac{\al_a}{2} \, \phi^a \, \bar\pa^2 \bar{\phi}^a
\Bigg](\bar{z}) +\sum_{i=1}^M \Bigg[ \frac{1+\tilde{\al}_i}{4}  \, \la^i \, \bar\pa
\bar{\la}^i -\frac{1-\tilde{\al}_i}{4}  \, \bar\pa \la^i \, \bar{\la}^i
\Bigg](\bar{z}).
\label{Tv}
\ee 
There is a primary spin-3  operator that is quadratic in the fundamental fields
\be
W_3^v(\bar{z})=\left(\bar{\pa}\phi^a\bar\pa^2\bar{\phi}^a-\bar\pa^2\phi^a\bar{\pa}\bar{\phi}^a+\frac{1}{2}\left(\bar{\pa}\bar{\lambda}^i\,\bar{\pa}\lambda^i-\bar\pa^2\bar{\lambda}^i\,\lambda^i\right)\right)(\bar{z})\ .
\ee
It has the following OPE
\bea
W_3^v(\bar{z})W_3^v(\bar{w})&\sim &\frac{c/3}{(\bar{z}-\bar{w})^6}+\frac{2 T(\bar{w})}{(\bar{z}-\bar{w})^4}
+\frac{\bar{\pa} T(\bar{w})}{(\bar{z}-\bar{w})^3}\\
&& +
\frac{\sqrt{\frac{512 (c+2)}{15 c+66}} W^v_4(\bar{w})-\frac{3}{10}\bar\pa^2 T(\bar{w})+\frac{32}{5 c+22}(TT-\frac{3}{10}\bar{\pa}^2 T)(\bar{w}) }{(\bar{z}-\bar{w})^2} \label{c334v}\\
&& +
\frac{\sqrt{\frac{512 (c+2)}{15 c+66}}\bar{\pa}W^v_4(\bar{w})-\frac{3}{10}\bar\pa^3 T(\bar{w})+\frac{32}{5 c+22}\bar{\pa}(TT-\frac{2}{15}\bar{\pa}^2 T)(\bar{w}) }{2(\bar{z}-\bar{w})}\,,\qquad 
\eea
where we have defined a primary spin-4 operator 
\bea
\nonumber W_4^v(\bar{z})&=&\sqrt{\frac{5c+22}{2400 (c+2)}}\left(2\bar{\pa}\phi^a\bar\pa^3\bar{\phi}^a-6\bar\pa^2\phi^a\bar\pa^2\bar{\phi}^a+2\bar\pa^3\phi^a\bar{\pa}\bar{\phi}^a-\bar{\pa}\bar{\lambda}^a\,\bar\pa^2\lambda^a\right.\\
&&\qquad \left.+3\bar\pa^2\bar{\lambda}^i\,\bar{\pa}\lambda^i-\bar\pa^3\bar{\lambda}^i\,\lambda^i -\frac{120}{5c+22}(TT-\frac{3}{10}\bar{\pa}^2 T) \right)(\bar{z})\,,
\eea
so that is has the normalization
\be
W_4^v(\bar{z})W_4^v(\bar{w}) \sim \frac{c/4}{(\bar{z}-\bar{w})^8}+\mathcal{O}((\bar{z}-\bar{w})^{-7})\ .
\ee
Notice that we call the higher-spin algebra generated by these operators to be quadratic in the sense that the single particle operators, 
namely the terms with only one sum of the $i$ or the $a$ indices, are all quadratic  in the fundamental fields. The last term, which involves the normal ordered product $TT$, in the $W^v_4$ operators is added so that this operator is written in the primary basis; if we had chosen to work in the equally well-defined quasi-primary basis, this normal ordered product term can be dropped and all the terms in the operator become manifestly quadratic in the fundamental fields. 

We have checked to a few more higher operators and we find only one operator at each spin. Therefore we conclude this is a conventional higher-spin type $\mathcal{W}_{\infty}$ algebra. In particular, from the OPE (\ref{c334v}) we read out the structure constant 
\be
\left(c_{334}^v\right)^2=  \frac{512 (c+2)}{3(5 c+22)}\,,
\ee
which means this algebra is isomorphic to the $\lambda =1 $ case of the family of $\mathcal{W}_{\infty}[\lambda]$ algebra~\cite{Hornfeck:1993kp,Blumenhagen:1994wg,Gaberdiel:2012ku}.

\subsection{The ``horizontal"  higher-spin subalgebra}\label{hsh}

Since the algebra discussed in the last section is quadratic in terms of the fundamental fields, it is easy to check that all other operators in the chiral algebra form various representations of the vertical higher-spin algebra. In particular, single-sum operators with different number of fundamental fields are in different representations. 

In the following we will discuss an interesting observation that there is a higher-spin algebra structure on this set of representations. To see this we consider the higher-spin primary state of each representation of the vertical higher-spin algebra, namely the states with the lowest weight in each representation. 
It turns out that these lowest weight operators, which consist of single-sum terms of higher powers of the fundamental fields, generate another higher-spin algebra. 

We now explicitly construct the first few higher-spin operators in the horizontal algebra.  
The spin-2 operator is again~(\ref{T}).  
The primary spin-$3$ operator of this second higher-spin algebra is
\bea
 W^h_3(\bar{z}) & = &\Bigg(\frac{4}{3} \sqrt{\frac{2}{3}} \phi ^a  \phi ^a  \phi ^a  \bar\pa\bar{\phi }^a  \bar\pa\bar{\phi }^a  \bar\pa\bar{\phi }^a
-2 \sqrt{6} \phi ^a  \phi ^a  \partial^2_{\bar{z}}\bar{\phi }^a  \bar\pa\bar{\phi }^a
+2 \sqrt{6}\bar\pa\phi ^a  \phi ^a  \bar\pa\bar{\phi }^a  \bar\pa\bar{\phi }^a\qquad\qquad \\
&&+2 \sqrt{\frac{2}{3}} \phi ^a  \bar\pa^3\bar{\phi }^a
+\sqrt{\frac{3}{2}} \bar\pa^2\phi ^a  \bar\pa\bar{\phi }^a
-5 \sqrt{\frac{3}{2}} \bar\pa\phi ^a  \bar\pa^2\bar{\phi }^a 
-\frac{ \bar\pa\bar{\lambda }^i  \bar\pa\lambda ^i}{2\sqrt{6}}
+\frac{ \bar\pa^2\bar{\lambda }^i  \lambda ^i}{2\sqrt{6}}\Bigg) (\bar{z})\ .\quad \nonumber
\eea
The Virasoro primary spin-4 operator reads
\bea
\nonumber W^h_4(\bar{z})&=& -\frac{\sqrt{\frac{5 c+22}{c+2}}}{10 \sqrt{6}} \Bigg(-20 \phi ^a   \phi ^a   \phi ^a   \phi ^a   \bar\pa\bar{\phi }^a   \bar\pa\bar{\phi }^a   \bar\pa\bar{\phi }^a   \bar\pa\bar{\phi }^a -140 \bar\pa \phi ^a  \phi ^a   \phi ^a   \bar\pa\bar{\phi }^a   \bar\pa\bar{\phi }^a   \bar\pa\bar{\phi }^a \\
\nonumber && +160 \phi ^a   \phi ^a   \phi ^a   \bar\pa^2 \bar{\phi} ^a  \bar\pa\bar{\phi }^a   \bar\pa\bar{\phi }^a-105 \phi ^a   \phi ^a   \bar\pa^2 \bar{\phi} ^a  \bar\pa^2 \bar{\phi} ^a+480 \bar\pa \phi ^a \phi ^a   \bar\pa^2 \bar{\phi} ^a  \bar\pa\bar{\phi }^a
\\
\nonumber &&-90 \phi ^a  \phi ^a  \bar\pa^3 \bar{\phi} ^a  \bar\pa\bar{\phi }^a-75 \bar\pa \phi ^a  \bar\pa \phi ^a  \bar\pa\bar{\phi }^a  \bar\pa\bar{\phi }^a-60 \partial^2 \phi ^a \phi ^a  \bar\pa\bar{\phi }^a  \bar\pa\bar{\phi }^a\\
\nonumber && +78 \partial^2 \phi ^a \bar\pa^2 \bar{\phi} ^a-106 \partial \phi ^a  \bar\pa^3 \bar{\phi} ^a-6 \partial^3 \phi ^a  \bar\pa\bar{\phi }^a+15 \phi ^a  \bar\pa^4 \bar{\phi} ^a+\frac{1}{2} \bar\pa\bar{\lambda }^i  \bar\pa^2\lambda ^i\\
&&-\frac{3}{2} \bar\pa^2\bar{\lambda }^i \bar\pa \lambda ^i+\frac{1}{2} \bar\pa^3\bar{\lambda }^i  \lambda ^i+\frac{60}{5 c+22} \left(TT-\frac{3}{10 } \bar\pa^2 T\right)\Bigg)(\bar{z})\,,
\eea
where the repeated indices are summed over.
The OPE between $W^h_3(\bar{z})$ and itself turns out to be
\bea
 W^h_3(\bar{z}) \, W^h_3(\bar{w}) & \sim &
\frac{c/3}{(\bar{z}-\bar{w})^6}  +\frac{2T(\bar{w})}{(\bar{z}-\bar{w})^4} \,
 + \frac{\bar{\pa} \, T(\bar{w})}{(\bar{z}-\bar{w})^3}  \,  \label{w3h} \\
\nonumber && +
\frac{\sqrt{\frac{512 (c+2)}{15 c+66}} W^h_4(\bar{w})-\frac{3}{10}\bar\pa^2 T(\bar{w})+\frac{32}{5 c+22}(TT-\frac{3}{10}\bar{\pa}^2 T)(\bar{w})  }{(\bar{z}-\bar{w})^2}  \qquad\qquad\\
&& +
\frac{\sqrt{\frac{512 (c+2)}{15 c+66}}\bar{\pa}W^h_4(\bar{w})-\frac{3}{10}\bar\pa^3 T(\bar{w})+\frac{32}{5 c+22}\bar{\pa}(TT-\frac{2}{15}\bar{\pa}^2 T)(\bar{w}) }{2(\bar{z}-\bar{w})}\ .\nonumber
\eea
As the vertical algebra we checked a few more operators with higher spins and we find the generators of this algebra consists of one operators for each spin. 
This result, together with the structure constant 
\be
\left(c_{334}^h\right)^2=  \frac{512 (c+2)}{3(5 c+22)}\,,
\ee
that can be read off from~(\ref{w3h}),   again indicates that this algebra is isomorphic to the $\lambda=1$ case of the $\mathcal{W}_{\infty}[\lambda]$ algebra.

\subsection{Higher spin square}\label{hss}

The results in the previous subsections remind us about a similar structure, namely the ``higher spin square", in 2d symmetric orbifold CFT \cite{Gaberdiel:2014cha, Gaberdiel:2015mra,Gaberdiel:2015wpo}. In that context there are two different higher-spin symmetry algebras. One, which is referred to as the vertical higher-spin algebra, is generated by a tower of higher-spin generators that are quadratic in the fundamental fields. The other algebra, referred to as the horizontal higher-spin algebras, are generated by higher-spin operators that are higher powers of the fundamental fields that subject to a single sum of the repeated indices. By taking the commutators of these two algebras, a larger chiral algebra is generated and the new generators can all be organized into representation of the vertical higher-spin algebra.

In our model, we find a very similar structure at the $q\to \frac{N}{M}$ limit. The algebra discussed in section \ref{hsv} correspond to the generators of the vertical higher-spin algebra, and hence they are labeled by upper indices $v$. 
On the other hand, the operator discussed in section \ref{hsh} is analogous to the horizontal higher-spin algebra, and hence they are labeled by upper indices $h$.  The OPE of the operators in the two higher-spin algebra is non-trivial and  one finds a much larger chiral algebra being generated. The latter is the realization of the stringy ``higher spin square" in our model. 

To see this explicitly, we can consider the OPE of the spin-3 operators from the two higher-spin algebras
\be
W_3^v(\bar{z})W_3^h(\bar{w})=\frac{6N+\frac{2}{3}M}{(\bar{z}-\bar{w})^6}+\frac{\frac{2(M+9N)}{N-M}T(\bar{w})-4W^{s,1}_2(\bar{w})}{(\bar{z}-\bar{w})^4}+\mathcal{O}\left((\bar{z}-\bar{w})^{-3}\right)\,,
\ee
where a new Virasoro primary spin-2 operator appears on the 4th order pole
\be
W^{s,1}_2(\bar{z})=\left(\frac{5 N }{2 (M-N)}\bar{\pa}\bar{\lambda }^i  \lambda ^i-\frac{5 N }{M-N}\bar{\pa}\phi ^a \bar{\pa} \bar{\phi }^a-4 \phi ^a  \bar{\pa}^2\bar{\phi }^a+3 \bar{\pa}\phi ^a \bar{\pa}\bar{\phi }^a+2 \phi ^a  \phi ^a  \bar{\pa}\bar{\phi }^a  \bar{\pa}\bar{\phi }^a\right)(\bar{z})\ .
\ee
It satisfies the following OPE
\bea
\nonumber W^{s,1}_2(\bar{z})W^{s,1}_2(\bar{w})&\sim& \frac{\frac{N (32 N-7 M)}{M-N}}{(\bar{z}-\bar{w})^4}+\frac{\frac{2 N (7 M-32 N)}{(M-N)^2} T(\bar{w})+ \frac{4 (M-6 N)}{M-N}W^{s,1}_2(\bar{w}) +\frac{3}{M-N} W^{s,2}_2(\bar{w}) }{(\bar{z}-\bar{w})^2}\quad \\
&&+\frac{\frac{ N (7 M-32 N)}{(M-N)^2} \bar{\pa} T(\bar{w})+ \frac{2 (M-6 N)}{M-N}\bar{\pa} W^{s,1}_2(\bar{w}) +\frac{3}{2(M-N)} \bar{\pa} W^{s,2}_2(\bar{w}) }{(\bar{z}-\bar{w})}\,,\qquad 
\eea
where a second new spin-2 field appears
\be
W^{s,2}_2(\bar{z})=\left(2 M \bar{\pa}\phi ^a  \bar{\pa}\bar{\phi }^a-N\bar{\pa}\bar{\lambda }^i  \lambda ^i\right)(\bar{z})\ .
\ee
The OPE among the 3 spin-2 fields closes among themselves
\bea
\nonumber W^{s,2}_2(\bar{z})W^{s,2}_2(\bar{w})&\sim& \frac{4MN{(M-N)}}{(\bar{z}-\bar{w})^4}+\frac{-8MN \, T(\bar{w}) +4{(M+N)} W^{s,2}_2(\bar{w}) }{(\bar{z}-\bar{w})^2}\\
&&+\frac{-4MN\, \bar{\pa}T(\bar{w})+ 2{(M+N)} \bar{\pa}W^{s,2}_2(\bar{w}) }{(\bar{z}-\bar{w})}\,,\\
\nonumber W^{s,1}_2(\bar{z})W^{s,2}_2(\bar{w})&\sim& \frac{-10MN}{(\bar{z}-\bar{w})^4}+\frac{\frac{20 M N }{M-N} T(\bar{w})+  4M\,W^{s,1}_2(\bar{w}) -\frac{10N}{M-N} W^{s,2}_2(\bar{w}) }{(\bar{z}-\bar{w})^2}\quad \\
&&+\frac{\frac{10 M N }{M-N} \bar{\pa}T(\bar{w})+ +2M\,\bar{\pa}W^{s,1}_2(\bar{w}) -\frac{5N}{M-N} \bar{\pa}W^{s,2}_2(\bar{w}) }{(\bar{z}-\bar{w})}\ .
\eea
Notice that the central terms of the $W^{s,1}_2W^{s,1}_2$ and $W^{s,2}_2W^{s,2}_2$ OPE are negative, and one would doubt if this means there are non-unitarity issue in this algebra. We think this is not the case. We think the negative norm is due to the impact of the spin-1 operator that we have not properly factored out, and probably also the fact that this higher spin square is only a subalgebra of the much larger chiral algebra. Notice we consider the higher-spin square as a subalgebra of the even larger chiral algebra. Although the operators in the higher spin square do close among themselves, they do not commute with the other operators of the chiral algebra that are not in the higher-spin square. The spin-1 field is one example; it has nontrivial commutation relations with most of the operators in the higher-spin square and the commutators contains operators that are not in the higher spin square. Given that the central terms of the stress energy tensor and the spin-1 fields are both positive, and the IR fixed point of the model (\ref{J1}) is unitary. We believe there is no issue of non-unitarity; once the spin-1 fields are properly factored out, the central terms of all fields will be positive definite. We have seen partial evidence of this statement, but to verify this statement explicitly we have to properly factor out the spin-1 fields for all the operators in the higher-spin square. We defer this into a future project.
  
From this result, we observe that although there are no  $M/N$ dependence in the two higher-spin algebras discussed in section~\ref{hsv}, \ref{hsh}, there are indeed explicitly $N/M$ or $q$ dependence in the higher-spin square. Since the two higher-spin algebras are fixed, both being the $\mathcal{W}_{\infty}[1]$ algebra, there is not likely further field redefinitions to remove the $q$ dependence. Therefore $q$ should be regarded as a new parameter that labels a family of different higher-spin square structure. 

\subsection{Relation with the  emergent IR higher-spin symmetry}\label{rel}

We have shown that there is not a single-sum $\mathcal{W}_{\infty}$ subalgebra in the chiral algebra of the $\bar{Q}$-cohomology at any value of $q\neq \frac{N}{M}$. 
On the other hand, in the limit $q\to \frac{N}{M}$, we found two different higher-spin subalgebra that further generate a higher-spin square subalgebra.  The $q\to \frac{N}{M}$ is the same limit where the IR higher-spin symmetry discussed in~\cite{Peng:2018zap} emerges.
Therefore it is natural to expect a close relation between the higher-spin algebra in the $\bar{Q}$-cohomology and  the emergent IR higher-spin symmetry. But one would like to ask which higher-spin subalgebra, among those discussed in the section~\ref{hsv},\ref{hsh} and \ref{hss}, is the ``image" of this IR higher-spin algebra. 

We believe that the vertical higher-spin algebra is the counterpart of the IR higher-spin algebra in the $\bar{Q}$-cohomology that can be extended away from the IR fixed point. The reason is the following. The emergent IR higher-spin symmetry is observed in the singlet channel of the 4-point function $\langle\bar\phi^a\phi^a \bar\phi^b\phi^b\rangle$ and $\langle\bar\phi^a\phi^a\bar\Lambda^i\Lambda^i\rangle$. The antiholomorphic higher-spin operators running in this channel have the schematic form 
\be
\mathcal{O}^s\sim \bar{\pa}\bar\Phi^a\bar{\pa}^{s-1}\Phi^a+\bar\Lambda^i\bar{\pa}^{s-1}\Lambda^i+c(H,\mathcal{O}^s)\,,\label{os}
\ee 
where the first two terms represent schematically the free field expressions of the higher-spin operators, and the last term represents all the terms generated from the evolution by the Hamiltonian of the model. Due to the presence of the last term, we do not expect simple expressions of these operators in terms of the fundamental fields $\Phi^a$ and $\Lambda^i$. One the other hand, as explained originally in~\cite{Witten:1993jg}, which we briefly recast in section~\ref{caq}, the operators in the $\bar{Q}$-cohomology and the algebraic relations among them  continue to make sense even if the potential is tune to zero. In this imaginary process the operators~(\ref{os}) approaches their free field expressions, which makes it clear that their counterparts in the  $\bar{Q}$-cohomology should be those operators that are quadratic in the fundamental fields. \footnote{This also means that the operator growth dressing, see e.g.~\cite{Roberts:2018mnp,Qi:2018bje} and the reference therein, to the operators  by the Hamiltonian evolution in the full theory is not observed in the chiral algebra. }
As a result, it is natural to consider the vertical higher-spin algebra found in section~\ref{hsv} to be related to the emergent IR higher-spin symmetry algebra discussed in~\cite{Peng:2018zap}.

\section{Conclusion}

In this paper we consider the chiral algebra of a class of 1+1 dimensional SYK models with $\mathcal{N}=(0,2)$ and $\mathcal{N}=(2,2)$ supersymmetry. 

In the special limit $q\to \frac{N}{M}$ we have constructed two different higher-spin algebras: one algebra is generated by bilinear ``single-sum" operators, the other is generated by ``single-sum" operators that are higher powers  in the fundamental fields. The two algebras then generate a larger stringy algebra that has the structure of a ``higher spin square". This gigantic stringy symmetry has been discussed previously \cite{Gaberdiel:2014cha, Gaberdiel:2015mra,Gaberdiel:2015wpo}. It can be used as a guiding symmetry to organize the spectrum of certain string theories, and it is useful to clarify the relation between string theory, in some appropriate tensionless limit, and higher-spin theories. It is not surprising that a similar structure appears   in the SYK model discussed in \cite{Peng:2018zap} and in this paper. The results in \cite{Peng:2018zap} indicates that as we tune a parameter, the model exhibits emergent higher-spin symmetry and mimics the transition from a tensile string theory to its tensionless limit. Therefore, the appearance of similar higher spin square structure becomes natural. As commented in section~\ref{rel}, it is natural to identify the vertical higher-spin subalgebra, which is generated by operators that are bilinear in the fundamental fields, with the emergent IR higher-spin algebra. 
It is thus very interesting to further clarify the explicit mapping between the two set of higher-spin-type symmetries.

\vspace{3mm}
\centerline{\bf Acknowledgments}
\vspace{1mm}

We thank Micha Berkooz, Zhen Bi, Chi-Ming Chang, Sean Colin-Ellerin, Johanna Erdmenger, Yingfei Gu, Antal Jevicki, Chaoming Jian, Xiaoliang Qi, Emil Martinec, Fidel Schaposnik Massolo, Jeff Murugan, Mukund Rangamani, Marcus Spradlin, Stefan Stanojevic, Herman Verlinde, Anastasia Volovich, Junggi Yoon and Pengfei Zhang for helpful discussions on related topics. We are especially grateful to Matthias Gaberdiel for valuable comments on the draft version of this paper.  
CA thanks the participants for their inputs in
the Friday Advanced Physics Graduate Seminar organized by Martin Rocek.
CP thanks the hospitality of the Aspen Center for Physics and the Kavili institute for Thoretical Physics during the various stages of this work. 
This research by CA was supported by Basic Science Research Program through the National Research Foundation of Korea funded by the Ministry of Education(No. 2017R1D1A1A09079512). The work of CP was supported by the US Department of Energy under contract DE-SC0010010 Task A.  This work was performed in part at Aspen Center for Physics, which is supported by National Science Foundation grant PHY-1607611. This research was also supported in part by the National Science Foundation under Grant No. NSF PHY-1748958.

\appendix

\renewcommand{\theequation}{\Alph{section}\mbox{.}\arabic{equation}}


\end{document}